\documentclass{amsart}

\usepackage{t1enc,newlfont}
\usepackage[latin5]{inputenc}
\usepackage[dvips]{graphicx}

\begin{document}
\title[Effects of Lagrangian Multipliers on SWCNT in Real Space]
{Effects of Lagrangian Multipliers on SWCNT in Real Space}

\author{N. S\"{u}nel$^1$, E. Rizao\u{g}lu$^2$, K. Harigaya$^3$ and O. \"{O}zsoy$^1$}

\address{$^1$Department of Physics, Faculty of Arts and
Sciences, Gaziosmanpa\c{s}a University, Tokat, 60240 Turkey\\
$^2$Department of Physics, Faculty of Sciences, Istanbul
University, Istanbul, 34459 Turkey\\ $^3$Nanotechnology Research
Institute, AIST, Tsukuba 305-8568 Japan}

\email{osozsoy@gmail.com}


\begin{abstract}

Electronic properties, band width, band gap and van Hove
singularities, of (3,0), (4,0) and (9,0) zigzag nanotubes are
comparatively investigated in the Harigaya's model and a toy model
including the contributions of bonds of different types to the SSH
Hamiltonian differently. Optical transition frequencies are
calculated. In this way an experimental correlation between the
two models is achieved.
\end{abstract}

\maketitle
\section{Introduction}

The last two decades have seen an explosive growth of the
investigations on nanoscale materials. Carbon nanotubes have taken
a rather large part of this effort by blazing a trail in
experimental and theoretical studies. This is due to their unique
chemical, electrical and mechanical properties because of the
one-dimensional (1D) confinement of their electronic states,
resulting in the so-called van Hove singularities (vHSs) in the
density of states (DOS). With their engrossing properties they
have enormous applications in very different areas, ranging from
electronics to biotechnology. Research on controlled synthesis of
nanotubes to progress the standards of applications is still a
grand challenge. Recent progresses may make single-wall carbon
nanotubes (SWCNTs) to play an important role in future nanometer
scale integrated photonic devices, in quantum optics and in
biological sensing [1].

SWCNTs can be thought of as graphite cylinders formed by
connecting two crystallographically equivalent sites in a graphite
sheet.  A graphite sheet is a $\pi$-system based on $sp^2$
hybridized carbon atoms consisting of strongly bound
$\sigma$-electrons and a zero-gap semiconductor in the sense that
the conduction and valance bands consisting $\pi$ states cross at
K and K$^{\prime}$ points of Brilloin zone. As derived from the
graphite sheet by the zone folding method, a ($n$, $m$) type tube
is metallic, possessing nonzero DOS in the region near Fermi
energy, if $n-m$ is a multiple of 3 since electronic band crosses
the Fermi level at two-thirds of the way to the zone edge. These
are correct as far as we remain true to zone folding approximation
[2].

The tight\,-\,binding approximation has played an important role
as one of the most commonly used calculation method for
quasi\,-\,one\,-\,dimensional and higher-dimensional systems. The
Su-Schrieffer-Heeger (SSH) model acquiesces this approximation and
is commonly used to study the stability and electronic structure
of nanotubes. In this model, hops of the $\pi$ electrons between
neighboring carbon atoms are studied and the local electron-phonon
interaction is treated in an adiabatic approximation [3,4].

The SSH model treats the roles of all bonds, tilt or right, on an
equal footing. However, chainlike structure of graphene and hence
of nanotubes brings in mind that the role of tilt and right bonds
in the lattice and electronic structure of nanotubes might be
different. As a matter of fact Cabria \emph{et al.} [5] during an
investigation of the stability of narrow zigzag carbon nanotubes
in the context of local density functional calculations found that
the axial bonds becoming shorter and more $sp^2$-like in
character, while the other bonds are becoming longer and more
$sp^3$-like in character.

Inspired by these, we found it worthwhile to modify the SSH
Hamiltonian used by Harigaya for studying the stability and energy
gap properties of nanotubes [6-10] by considering the
contributions of bonds of different types to the SSH Hamiltonian
differently at the very beginning. We introduced this modified
version of Harigaya's model and its application to armchair type
nanotubes in [11] and some other preliminary results elsewhere
[12]. Concerning the (3,0) zigzag nanotube, we observed that the
tiny energy gap appearing in Harigaya's model was lost, a result
consistent with the metallicity of this tube [13]. In this work,
we continue to investigate the electronic structure of zigzag type
nanotubes.

\section{Model and Calculation}

The aforementioned modified version of Harigaya's model is based
on the SSH Hamiltonian which includes different roles of the tilt
and right bonds. This is achieved by taking the hopping integral
of the ideal undimerized system, the electron-phonon coupling
constants and the spring constants different for the tilt and
right bonds. In this way the introduction of two Lagrange
multipliers and hence two constraints become unavoidable. The
separation of the constraint into two constraints, vanishing of
the sum of right bond distortions and vanishing of the sum of tilt
bond distortions, provides more freedom for lattice relaxations.

The total energy of the system is read as

\begin{equation}
E_{\mathrm T} = \sum_{i,\sigma}^{\prime} \epsilon_{i,\sigma} +
\frac{1}{2\gamma^{{\mathrm{t}}}}\sum_{i}[v_i^{{\mathrm{t}}}]^2 +
\frac{1}{2\gamma^{{\mathrm{r}}}}\sum_{i}[v_i^{{\mathrm{r}}}]^2\,,
\end{equation}

where $\epsilon_{j,\sigma}$s are the eigenvalues of the model
Hamiltonian. The self\,-\,consistent equation for the lattice is
extracted as

\begin{equation}
v_i^{{\mathrm{t,r}}}=2\gamma^{{\mathrm{t,r}}}\Big[\frac{\alpha^{{\mathrm{t,r}}}
C^{{\mathrm{t,r}}}}{\gamma^{{\mathrm{t,r}}}} -\frac{1}{N_{\mathrm
b}^{{\mathrm{t,r}}}}\frac{\gamma^{{\mathrm{t,r}}}}{\alpha^{{\mathrm{t,r}}}}\sum_{j,\sigma}^{\prime}B^{\dag}_{i+
1,j,\sigma}B_{i,j,\sigma}\Big]\,.
\end{equation}

Above $\mathbf{B}$s are the eigenvectors of the Hamiltonian and
$\gamma^{{\mathrm{t,r}}}=[\alpha^{{\mathrm{t,r}}}]^2/\kappa^{{\mathrm{t,r}}}$.
The prime indicates the sum over the occupied states of electrons
and $N_{\mathrm b}$ is the total number of $\pi$\,-\,bonds. It is
1.5 times of the number of sites $N$. $E_{\mathrm T}$ and
$v^{\mathrm {t,r}}_i$'s are numerically calculated in the
iteration method by the adiabatic approximation for phonons and
the calculation was kept further to have converged once
$v^{\mathrm {t,r}}_i$ varied less than $10^{-5}$.

\section{Results and Discussions}

The determination of energy eigenvalues, $\epsilon_{i,\sigma}$s
and energy eigenstates, $\mathbf{B}$s by solving Schrodinger's
equations necessitates the knowledge of the matrix representation
of model's Hamiltonian. It is quite clear that this matrix
representation depends strictly on the geometry, \emph{i.e.},
whether one plans to do the calculations for a carbon sheet, for
an open ended nanotube or a nanotube with periodic boundaries,
since the Hamiltonian is different for different geometries.

To be able to calculate the electronic band structure properties
of different geometries all together we preferred to write the
matrix representation of electronic part of our Hamiltonian in a
compact manner as follows

\begin{equation}
\mathcal{H}_{\mathrm{SSH}}^{(el)}=\left (
\begin{tabular}{cccccc}
 $\mathcal{M}_A$     &$\mathcal{M}_B$   &$\mathcal{O}$     &.               &.             &$\mathcal{M}_D$ \\
 $\mathcal{M}_B$     &$\mathcal{M}_A$   &$\mathcal{M}_B$   &.               &.             &. \\
 $\mathcal{O}$       &$\mathcal{M}_B$   &$\mathcal{M}_A$   &+               &.             &. \\
 .                   &.                 &+                 &+               &+             &. \\
 .                   &.                 &.                 &+               &+             &+ \\
 $\mathcal{M}_D$     &.                 &.                 &.               &+             &$\mathcal{M}_A$ \\
\end{tabular} \right )
\end{equation}

where "$+$"s represents continuation of $\mathcal{M}_A$ and
$\mathcal{M}_B$ matrices and

$$
 \mathcal{M}_A=\left (
\begin{tabular}{cccccc}
  0            &$A$          &0            &.            &.               &$\beta_N C$     \\
  $A$          &0            &$A$          &.            &.               &.               \\
  0            &$A$          &.            &.            &.               &.               \\
  .            &.            &.            &.            &$A$             &0               \\
  .            &.            &.            &$A$          &0               &$A$             \\
  $\beta_N C$  &.            &.            &0            &$A$             &0               \\
\end{tabular} \right )\,,
$$

$$
\mathcal{M}_D=\left (
\begin{tabular}{cccccc}
 0             &.               &.               &.               &.             &. \\
 .             &$\beta_S B$     &.               &.               &.             &. \\
 .             &.               &0               &.               &.             &. \\
 .             &.               &.               &$\beta_S B$     &.             &. \\
 .             &.               &.               &.               &.             &. \\
 .            &.               &.               &.               &.              &$\beta_S B$ \\
\end{tabular} \right )\,,
$$

$$
\mathcal{M}_D=\left (
\begin{tabular}{cccccc}
  $\beta_T D$  &.   &.            &.  &.  &.             \\
  .            &0   &.            &.  &.  &.             \\
  .            &.   &$\beta_T D$  &.  &.  &.             \\
  .            &.   &.            &.  &.  &.             \\
  .            &.   &.            &.  &0  &.             \\
  .            &.   &.            &.  &.  &$\beta_T D$   \\
\end{tabular} \right )
$$
and

$$
\mathcal{O}=\left (
\begin{tabular}{cccccc}
 0             &0               &0               &.               &.             &0 \\
 0             &0               &0               &.               &.             &0 \\
 0             &0               &0               &.               &.             &0 \\
 .             &.               &.               &                &.             &. \\
 .             &.               &.               &.               &.             &. \\
 0             &.               &.               &.               &.             &0 \\
\end{tabular} \right )\,.
$$

Here, $A$, $B$, $C$ and $D$ stand for the hopping integrals
between the nearest\,-\,neighbor interaction sites in the same row
(tilt bonds); between the nearest\,-\,neighbor interaction sites
in different rows (right bonds), which causes the sheet structure;
between the first and last sites in the same row, which creates
zigzag nanotube, and between the first and the last rows, which
create the periodic boundaries, respectively. All the remaining
terms indicated by dots are zero. $\beta_{\mathrm S}$,
$\beta_{\mathrm N}$ and $\beta_{\mathrm T}$ represent the
evolution parameters for carbon sheet, open ended nanotube and
nanotube with periodic boundaries. Each beta parameter varies from
0 to 1. In the case of armchair nanotubes the $B$s and $D$s
exchange.

Because of our toy model gives a better result concerning the
electronic band structure of (3,0) nanotube, compared to the
Harigaya's model [13], we want to check its results from the point
of view of the other electronic structure properties of this
nanotube and also from the point of view of electronic structure
properties of a large radius conducting (9,0) nanotube together
with a semiconducting (4,0) nanotube. We perform the calculations
with $N = 2n$ and $K$ up to 100 for all tubes. For (3,0), the
total number of C-sites increases up to 600 for $K = 100$ and the
numbers of tilt and right bonds reach up to 600 and 297. These
numbers are 800, 800 and 396 for (4,0) while they are 1800, 1800
and 891 for (9,0). In the periodic boundaries case, the numbers of
right bonds keep themselves while the numbers of tilt bonds
increase 3, 4 and 9 more, respectively. Of course, it should be
mentioned that for zigzag tubes the integer $n$ is related to the
length while the integer $K$ is related to the diameter of tube.
We chose the numerical set of $t_0^{\mathrm {t,r}}$,
$\alpha^{\mathrm{t,r}}$ and $\kappa^{\mathrm{t,r}}$ parameters as
2.5 eV, 6.31 eV/\AA\, and 49.7 eV/\AA$^2$, respectively [10].

The energy eigenvalues when (3,0), (4,0) and (9,0) nanostructures
(graphene, open ended nanotubes and nanotubes possessing periodic
boundaries) evolve from trans-PA chains, both in the framework of
our toy model and in that of Harigaya's model, are obtained
numerically by taking $t_0$ and $\alpha$ values the same for tilt
and right bonds. From these eigenvalues we extracted the band
width and band gap variations together with the one-dimensional
vHSs which correspond to logarithmic discontinuities in DOS.

First, we report band width, \emph{i.e.,} the difference between
the highest and lowest energy eigenvalues. Figs.\,1(a), (b) and
(c) show band width for three systems comparatively in both
models. For (3,0) the band width  naturally starts with 8.73 eV,
the band width value of trans-PA chain having the suitable length.
According to the Harigaya's model, at the early stages of carbon
sheet formation band width rises steeply and it then enters a
sequence of rising and staying constant stages with small periods.
But, according to our toy model it regularly increases during the
whole sheet formation process. However, at the end of sheet
formation process, band width reaches the same value, 13.80 eV for
both models. During the completion of open ended nanotube both
models give the same behavior for band width. It reaches 15 eV.
According to both models, the band width stays almost constant at
the formation stage of nanotube with periodic boundary conditions.
If only the final value of band width would be considered, there
will not be any difference between the two models. For (4,0) the
band width starts with 9.16 eV, again the band width of trans-PA
of length yielding the (4,0) structures, and shows almost the same
behavior as the band width of (3,0). For (9,0) the same appears
with 9.75 eV, the band with of trans-PA of this length.

\begin{figure}[ht]
\begin{center}
\includegraphics[angle=0, width=12.0
cm]{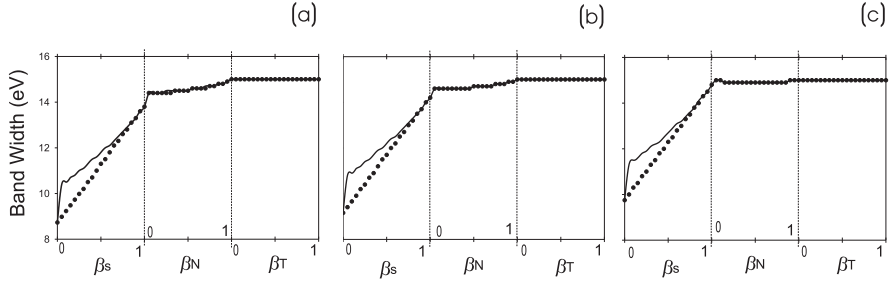} \caption{Band with evolution of (3,0), (4,0) and
(9,0) structures both in the Harigaya's model and in the toy
model. As $\beta_{\mathrm S}$ varies both $\beta_{\mathrm N}$ and
$\beta_{\mathrm T}$  are kept fixed at zero; as $\beta_{\mathrm
N}$ varies $\beta_{\mathrm S}$ is kept fixed at 1 and
$\beta_{\mathrm T}$ is kept fixed at zero and as $\beta_{\mathrm
T}$ varies both $\beta_{\mathrm S}$ and $\beta_{\mathrm T}$ are
kept fixed at 1. Solid and dotted lines correspond to the
variations in the Harigaya's model and in the toy model,
respectively.}
\end{center}
\end{figure}

Next, we show band gap, \emph{i.e.,} the difference between HOMO
and LUMO. For (3,0) the gap initiate with 3.39 eV, the gap value
of trans-PA. As graphene evolves, the gap goes down almost in a
parallel way for both models to 0. This is consistent of graphene
being a zero-gap semiconductor whose electronic structure near the
Fermi energy is given by an occupied $\pi$ band and an empty
$\pi^\ast$ band. It is quite clear from Fig.\,2(a) that in the
Harigaya's model the gap drops to zero for $t_0 = 1.63$ eV while
it drops to zero for $t_0^{\mathrm {t,r}} = 1.38$ eV in the toy
model. This is due to the looseness of the constraint conditions
compared to those in Harigaya's model. At the latest stages of
graphene formation and during the formation of open ended
nanotube, the gap vanishes and with the beginning of nanotube with
periodic boundaries it jumps up to 0.39 eV and falls to 0.33 eV at
the completion when Harigaya's model is used. Hence, according to
this model we have a nonzero gap for a (3,0) nanotube. SSH model
does not include bond angle and bond length variations giving rise
to finite curvature effects in nanotubes, detected in several
experiments as small energy gap near the Fermi level [14].
Therefore, the tiny gap near the Fermi level in (3,0) appearing
according to Harigaya's model can not be interpreted as due to
curvature effects. For this an extended form of tight binding
calculations to include curvature effects had to be used [15]. But
according to our toy model, the gap starts and ends with 0.08 eV
and 0.07 eV during the formation stage of (3,0) nanotube with
periodic boundary conditions (Fig.\,2(a)). For (4,0) the initial
gap value is 2.89 eV, it follows almost the same decrease up to
zero and then with the beginning of nanotube with periodic
boundaries jumps to relatively close values, 2.23 eV in the
Harigaya's model and 2.07 in the toy model (Fig.\,2(b)). Peng
\emph{et al.} [16] assume that this SWCNT is a semiconductor,
which was predicted to have an energy gap of about 2.5 eV based on
the graphene sheet model. On the other hand, some authors [5]
claim that (4,0) is metallic and this discrepancy arises from the
neglect of curvature effects in the graphene sheet model. As for
(9,0), the gap starts from an even smaller value, 1.91 eV and then
falls to zero at $t_0 = 0.88$ eV and at $t_0 = 75$ eV,
respectively. At the completion of nanotube with periodic
boundaries the gap goes to zero (Fig.\,2(c)).

\begin{figure}[ht]
\begin{center}
\includegraphics[angle=0, width=12.0
cm]{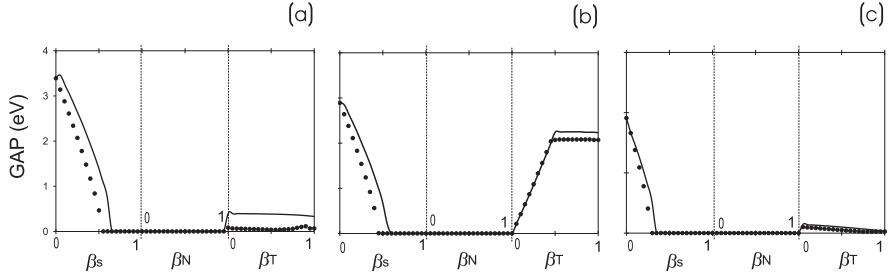} \caption{Band gap evolution of (3,0), (4,0) and
(9,0) structures both in the Harigaya's model and in the toy
model. As $\beta_{\mathrm S}$ varies both $\beta_{\mathrm N}$ and
$\beta_{\mathrm T}$  are kept fixed at zero; as $\beta_{\mathrm
N}$ varies $\beta_{\mathrm S}$ is kept fixed at 1 and
$\beta_{\mathrm T}$ is kept fixed at zero and as $\beta_{\mathrm
T}$ varies both $\beta_{\mathrm S}$ and $\beta_{\mathrm T}$ are
kept fixed at 1. Solid and dotted lines correspond to the
variations in the Harigaya's model and in the toy model,
respectively.}
\end{center}
\end{figure}

We will report on calculations of one-dimensional van Hove
singularities, \emph{i.e.,} logarithmic singularities: These
singularities may be important in experimental correlations of
models of nanotubes because optical absorption or emission rates
in nanotubes is related  primarily to the electronic states at the
vHSs [17]. In this respect, the electronic transition energies
$E_{ii}, \, i = 1, 2, 3, \ldots$, that is the energy separations
between $i^{\mathrm th}$ vHSs in the DOS of the conduction and
valence bands play the fundamental role since the measured
absorption peaks occur at these transition energies [18,19].

The recent progress in measuring the absorption peaks of single
SWCNT instead of SWCNTs bundle is encouraging [1]. In this way
effects arising from ensemble averaging and effects introduced by
a composite medium are eliminated. Besides this, combined
fluorescence and Raman scattering measurements on the single
molecule level open up new ways to characterize SWCNTs [1]. These
are among nondestructive characterization techniques of SWCNTs,
the only handicap is the effects on electronic properties due to
inhomogeneities in their local environments.

\begin{figure}[ht]
\begin{center}
\includegraphics[angle=0, width=12.0
cm]{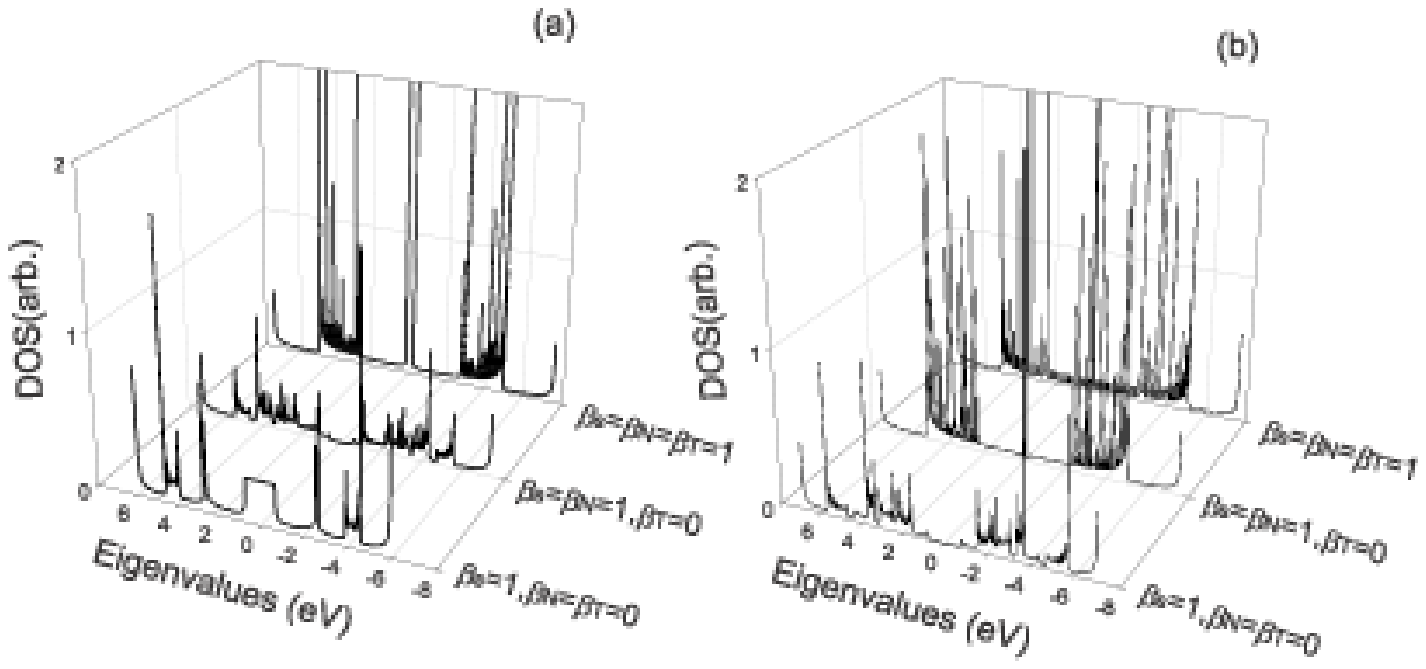} \caption{DOS evolution of (3,0 ) structures for K =
60. (a) for the Harigaya's model and (b) for the toy model.}
\end{center}
\end{figure}

\begin{figure}[ht]
\begin{center}
\includegraphics[angle=0, width=12.0
cm]{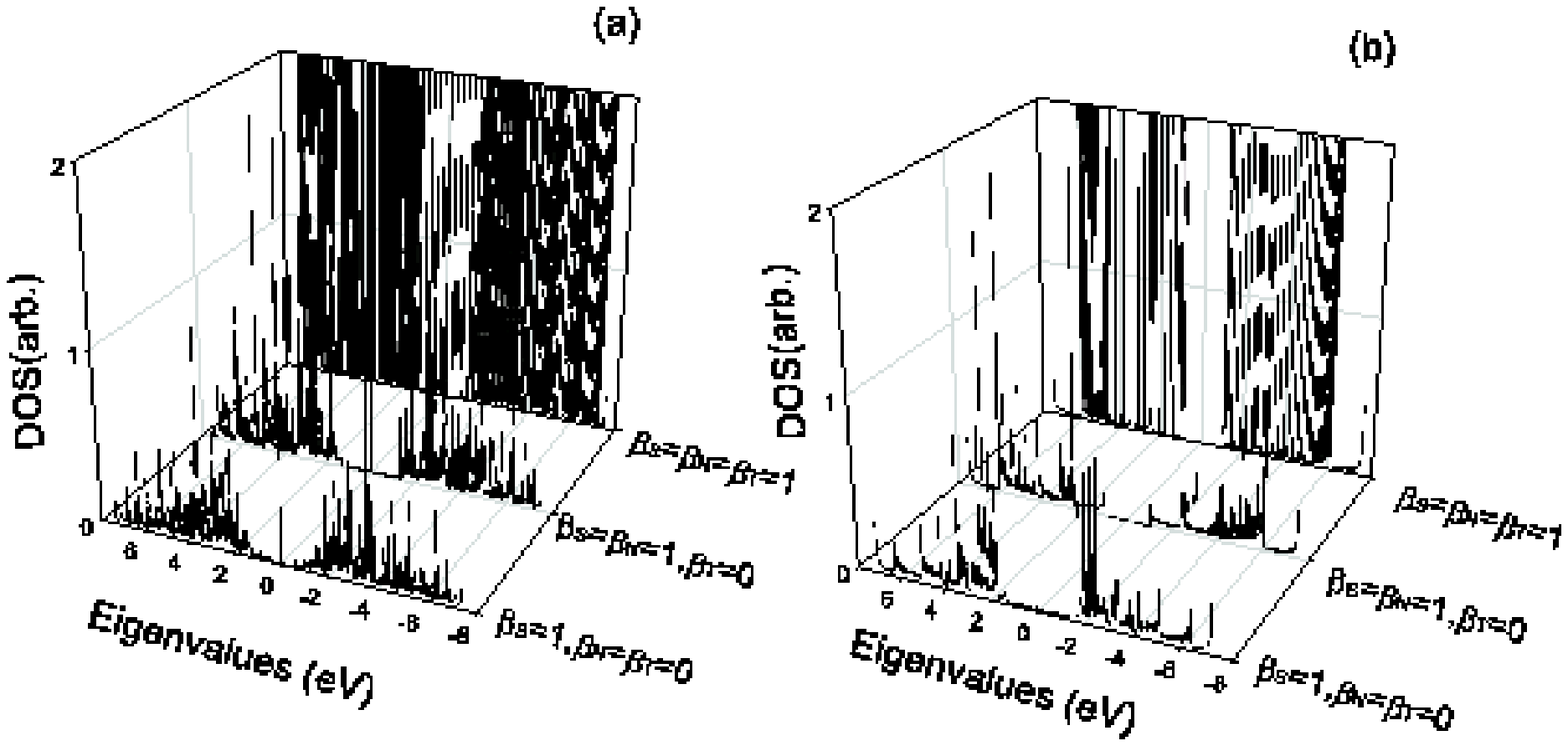} \caption{Same as Fig.3 but for (4,0 ).}
\end{center}
\end{figure}

\begin{figure}[ht]
\begin{center}
\includegraphics[angle=0, width=12.0
cm]{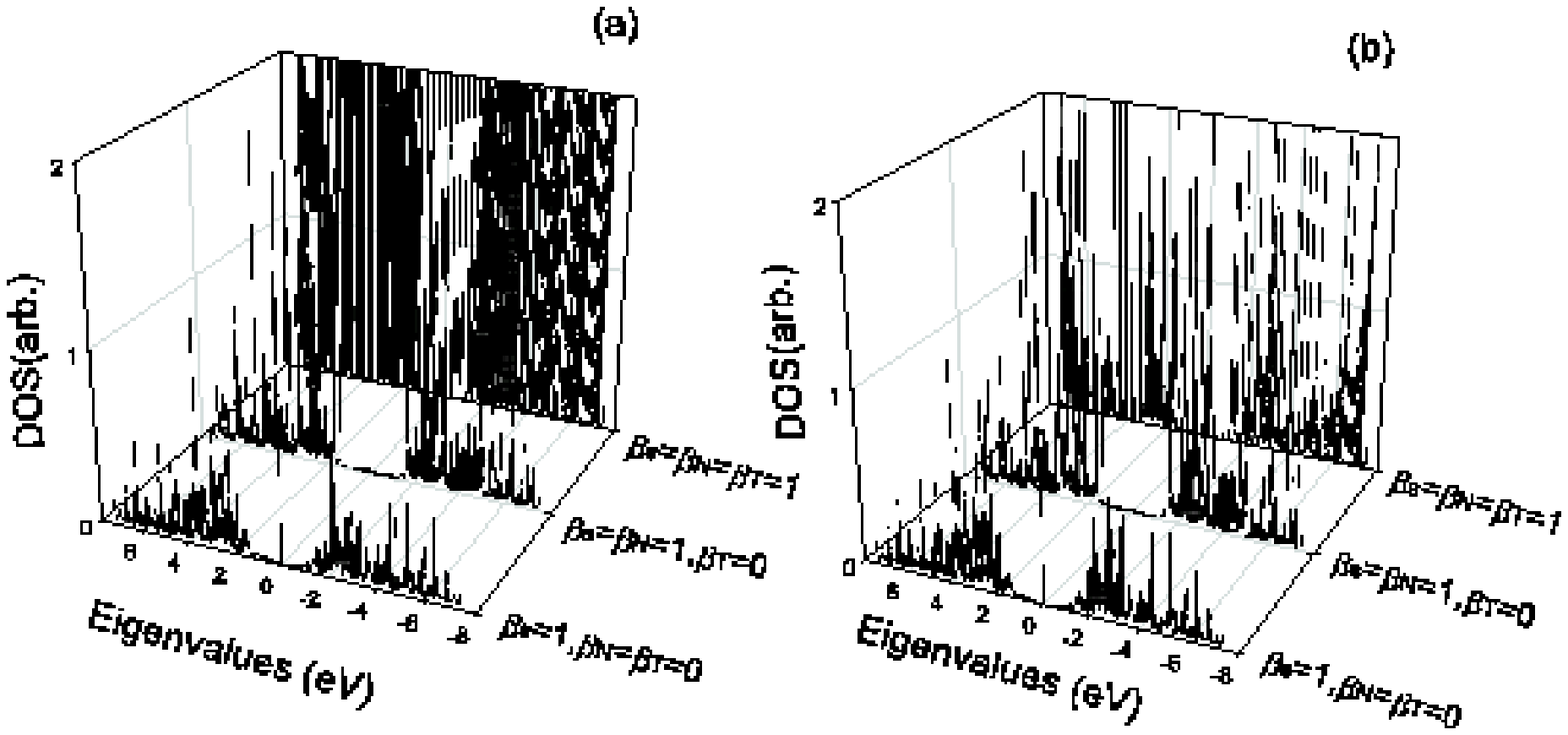} \caption{Same as Fig.3 and Fig.4 but for (9,0 ).}
\end{center}
\end{figure}

Sharp vHSs occur in carbon nanotubes with diameters less than 2 nm
[6]. All of our three SWCNTs have this property, 0.235, 0.313 nm
and 0.704 nm, respectively. Figs.\,3, 4 and 5 respectively depict
the DOS of graphene, open ended SWCNTs and nanotubes possessing
periodic boundaries according to the Harigaya's model and the toy
model for (3,0), (4,0) and (9,0) structures. We calculated the
energies at which these singularities appear. In the toy model the
number of vHSs is higher. Gap is the interval between the first
vHSs in the valance and conduction bands. For (3,0), the gap value
obtained in this way is consistent with the gap value obtained
from Fig.\, 2(b) both in the Harigaya's model and the toy model.
(4,0) has an energy gap of about 2.5 eV based on the graphene
sheet model [16]. Our result is consistent with this. For (9,0),
we obtain an extra van Hove singularity very close to Fermi energy
according to both models. This is reasonable since (9,0) nanotubes
are conducting. The next three singularities are coinciding with
those of Hartschuh \emph{et al.} [1].

\begin{table*}

\begin{tabular}{|c|c|c|c|c|c|c|c|c|c|c|}
\hline Tubes   &     \multicolumn{1}{|c|}{Models} &
\multicolumn{1}{|c|}{$\beta$s}  &

\multicolumn{6}{|c|}{   Frequencies (cm$^{-1}$)}                 \\
\hline

      & Harigaya   &  $\beta_{N}=1$   & 25249.3  &&&&& \\
(3,0) &    "       &  $\beta_{T}=1$   & 22168.4  & 23887.7 &&&&  \\
      & Toy        &  $\beta_{N}=1$   & 19988.1  & 26858.3 &&&& \\
      &    "       &  $\beta_{T}=1$   & 20910.7  & 23083.6 & 26941.8 &&& \\

\hline

      & Harigaya   &  $\beta_{N}=1$   & 19881.8  & 24102.7 & 28094 &&& \\
(4,0) &    "       &  $\beta_{T}=1$   & 19898    &&&&& \\
      & Toy        &  $\beta_{N}=1$   & 20285.1  & 24981.5 &&&& \\
      &    "       &  $\beta_{T}=1$   & 20285.1  &&&&& \\

\hline

      & Harigaya   &  $\beta_{N}=1$   & 14662     & 20669.7   & 25840 &&& \\
(9,0) &     "      &  $\beta_{T}=1$   & 11263.1   & 13193.4   & 18246.8 & 27226.6 &&  \\
      & Toy        &  $\beta_{N}=1$   & 10989.6   & 13350     & 13974.8 & 19703.3 & 24117.3 & 25765.3   \\
      &     "      &  $\beta_{T}=1$   & 11123.5   & 13318.4   & 18508.1 & 22206.5 & 23535.6 & 25611.1   \\

\hline
\end{tabular}

\caption{\label{tab:table1}Calculated optical transition
frequencies in both models for open ended   SWCNTs and SWCNTs with
periodic conditions. For the $\beta_{N}=1$ lines $\beta_{S}=1$ and
$\beta_{T}=0$ and for the lines $\beta_{T}=1$ we have
$\beta_{S}=\beta_{N}=1$.}

  \end{table*}

Transition frequencies $\omega_{ii} = E_{ii}/ \hbar$  cm$^{-1}$ in
the optical spectral range for (3,0), (4,0) and (9,0) open ended
nanotubes and nanotubes with periodic boundaries are selected from
the calculated transition energies and are shown in Table 1.

\section{Conclusion }

Electronic properties, band with; band gap and vHSs of three
zigzag nanotubes, (3,0) (small radius-conducting), (4,0)
(semiconducting) and (9,0) (large radius-conducting) are
comparatively studied in a toy model and in the Harigaya's model.
The toy model is an extension of the Harigaya's model, it includes
the contributions of bonds of different types to the SSH
Hamiltonian differently. Both models give the same band width. In
the (3,0) case, the band gap appearing according to the Harigaya's
model disappears when the toy model is used for periodic
conditions. Both models give the same band gap for (4,0) nanotube.
It agrees well with the gap value given in the literature. For
(9,0) the results of both models are the same. When the toy model
is used one gets many more vHSs. For (9,0) we obtained an extra
singularity in the vicinity of Fermi energy. The calculated next
three singularities agree with those given in the literature. We
could not find any data in the literature for the correlation of
the other singularities.

\section*{References}

[1] Hartschuh, A.;  Pedrosa, H.N.; Peterson, J.; Huang, L.; Anger,
P.; Qian, H.; Meixner, A.J.; Steiner, M.; Novotny, L.; Krauss,
T.D. Chem Phys Chem 2005, 6, 1.

[2] Dresselhaus, M.S.; Eklund, P.C. Advances in Phys 2000, 49,
705.

[3] Su, W.P.; Schrieffer, J.R.; Heeger, A.J. Phys Rev 1980, B22
2099.

[4] Stafstrom, S.; Riklund, R.; Chao, K.A. Phys Rev 1982, B26
4691.

[5] Cabria, I.; Mintmire, J.W.; White, C.T. Int J Quant Chem 2003,
91, 51.

[6] Harigaya, K. J Phys Soc Jpn 1991, 60, 4001.

[7] Harigaya, K. Phys Rev 1992, B45, 12071.

[8] Harigaya, K. Phys Rev 1999, B60, 1452.

[9] Harigaya, K. Synt Met 2003, 135, 751.

[10] Harigaya, K. J Phys: Condens Matter 1998, 10, 6845.

[11] S\"{u}nel, N.; Rizao\u{g}lu, E.; Harigaya, K.; \"{O}zsoy, O. Phys Lett
2005, A338, 366.

[12] S\"{u}nel, N.; \"{O}zsoy, O. Int J Quant Chem 2004, 100, 231.

[13] \"{O}zsoy, O.; S\"{u}nel, N. Czech J Phys 2004, 54, 1495.

[14] Itkis, M.E.; Nigoyi, S.; Meng, M.E.; Hamon, M.A.; Hu, H.;
Haddon, R.C. Nano Lett 2002, 2, 155.

[15] Dresselhaus, M.S.; Saito, R.; Jorio, A. Annual Review of
Materials Research; August 2004; Vol. 34; Pages 247-278.

[16] Peng, L.M.; Zhang, Z.L.; Xue, Z.Q.; Wu, Q.D.; Gu, Z.N.;
Pettifor, D.G. Phys Rev Lett 2000, 85, 3249.

[17] Cronin, S.B.; Swan, A.K.; \"{U}nl\"{u}, M.S.; Goldberg, B.B.;
Dresselhaus, M.S.; Tinkham, M. Phys Rev 2005, B72, 035425-1.

[18] Collins, J.E.; Sippel, J.; Arnason, S.; Rinzler, A.G.
University of Florida, Department of Physics, P.O. Box 118440,
Gainsesville, FL 32611-8440 U.S.A.

[19] Fantini, C.; Jorio, A.; Souza, M.; Strano, M.S.; Dresselhaus,
M.S.; Pimenta, M.A. Phys Rev Lett 2004, 93, 147406-1.

\end{document}